\documentclass[usenatbib]{mn2e}
\usepackage{tabularx,hhline,amsmath,amssymb}
\usepackage{graphics}
\usepackage[small]{subfigure}

\title[Luminosity in Bulges and Disks]{Quantitative Morphology 
of Galaxies from the SDSS I: Luminosity in Bulges and Disks}

\author[Lidia A.M. Tasca and Simon D.M. White]
       {Lidia A.M. Tasca$^1$\thanks{E-mail: lidia.tasca@oamp.fr} and Simon D.M. White$^1$\\
        $^1$Max-Planck-Institut f\"ur Astrophysik,
            Karl-Schwarzschild-Str. 1 85741 Garching b. M\"unchen, Germany}
\date{\today}

\pagerange{\pageref{firstpage}--\pageref{lastpage}}
\pubyear{2005}

\begin{document}

\maketitle
\label{firstpage} 
\begin{abstract}
In the first paper of this series we use the publicly available code {\bf
Gim2D} to model the $r$- and $i$-band images of all galaxies in a
magnitude-limited sample of roughly 1800 morphologically classified galaxies
taken from the Sloan Digital Sky Survey. The model is a concentric
superposition of two components, each with elliptical isophotes with constant
flattening and position angle. The disk luminosity profile is assumed
exponential, while the bulge is assumed to have a de Vaucouleurs or a S\'{e}rsic
profile. We find that the parameters returned by {\bf Gim2D} depend little on
the waveband or bulge profile used; their formal uncertainties are usually
small. Nevertheless, for bright galaxies the measured distribution of $b/a$,
the apparent disk flattening, deviates strongly from the expected uniform
distribution, showing that the ``disk'' identified by the code frequently
corresponds to an intrinsically 3-dimensional structure rather than to a true
thin disk.  We correct approximately for this systematic problem using the
observed statistics of the $b/a$ distribution and estimate, as a function of
absolute magnitude, the mean fractions of galaxy light in disks and in ``pure
bulge'' systems (those with no detectable disk). For the brightest galaxies
the disk light fraction is about 10\% and about 80\% are ``pure bulge''
systems. For faint galaxies most of the light is in disks and we do not detect
a ``pure bulge'' population. Averaging over the galaxy population as a whole,
we find that $54\pm 2\%$ of the local cosmic luminosity density at
both
$r$ and $i$ comes from
disks and $32\pm 2\%$ from ``pure bulge'' systems. The remaining $14\pm 2\%$
comes from bulges in galaxies with detectable disks.
\end{abstract}

\begin{keywords}
Survey -- galaxies: photometry -- galaxies: bulge--to--disk decomposition
-- galaxies: fundamental parameters -- 
galaxies: spirals -- galaxies: ellipticals -- galaxies: bulge
-- galaxies: disk 
\end{keywords}

\section{Introduction}
\label{sec:intro} 

Observations show that galaxies typically have two components with different
photometric and dynamical properties: a thin, rotationally supported stellar
disk which often also contains gas and dust, and a bulge, a spheroidal or
ellipsoidal component made purely of stars.  Most galaxies possess both
structures, but some, the ellipticals, have no significant disk, while others,
late--type spirals and irregulars, have little or no bulge.  Since these two
components presumably had different formation paths, their relative importance
must be a fundamental clue to how each galaxy formed.  Among the disk galaxies
themselves there exists a dichotomy between those rich in gas, dust and the
accompanying star formation, spiral and irregular galaxies, and those in
which this activity is virtually absent, lenticular galaxies.

It is known that galaxies of differing morphology are segregated according to
environmental density \citep{Dressler:80} with the low--density field composed
largely of spirals and irregulars and the densest regions of clusters composed
of lenticulars and ellipticals.  Many interpretations have been proposed for
this observation, ranging from initial condition biases which imprint the
differences at birth, to gravitational, gas dynamical or radiative processes
through which galactic environment affects later evolution. Recent studies
using the large, complete samples provided by the Sloan Digital Sky Survey
(SDSS, see below) show the distribution of galaxy mass to depend quite
strongly on environment, apparently reflecting an initial condition bias,
while for galaxies of given mass, star formation properties depend much more
strongly on environment than do structural properties, suggesting that
external processes primarily affect the gas component from which stars form
\citep{Kauffmann:04}.

Understanding the chronology of bulge and disk formation by analysing the
relative contributions of these two components at different cosmological
epochs is a fundamental goal of observational cosmology. Such data provide
important constraints on competing scenarios of galaxy formation and
evolution.  In models where bulges form first and disks are added later no
close correlation is expected between the two formation phases; the ratio
of spheroid luminosity to total luminosity measures the the efficiency of the
first burst of star formation relative to later slow accretion (Fall \&
Efstathiou 1980; Kauffmann, White \& Guiderdoni 1993). If instead bulges form
out of disks by secular evolution, then a stronger correlation between their
properties may be expected (Norman et al.1996; Bournaud \& Combes 2004).

The relative contributions of bulges and disks to the stellar content of
galaxies of differing mass and also to the overall stellar content of the
Universe are clearly important quantities which should be reproduced within
any viable picture of galaxy formation. It has long been known that the
massive galaxy population is dominated by light from spheroids, while disks
tend to dominate in lower mass populations (e.g. \citet{Efstathiou:82}), 
but there have
been no recent quantitative studies for the galaxy population as a whole. The
most commonly cited estimates of the fractions of all stars in bulges and
disks trace back to the work of \citet{Simien:86} and \citet{Schechter:87}.
These results suffer from poor statistics and are based largely on
photographic material and on visual ``decomposition'' of the light of each
galaxy into its bulge and disk fractions. With large photometric surveys,
linear CCD detectors and quantitative decomposition techniques it should
clearly be possible to do much better.  This is the goal of the present paper
and subsequent papers in this series.

We apply a modern two-dimensional morphological decomposition algorithm to the
images of a magnitude-limited sample of relatively nearby and bright galaxies
with photometry available from SDSS. In this paper we use the results to study
the overall contribution of disks and bulges to the light of the nearby galaxy
population both as a function of intrinsic galaxy luminosity and for the
population as a whole. In section \ref{sec:cap1} we describe the selection of
the sample we use. In section \ref{sec:cap2} we describe the decomposition
algorithm and the parametric functions it uses to fit the galaxy images.  We
also describe how the data were processed.  In section \ref{sec:cap3} results
are compared for different photometric bands and for different parametric
fitting functions in order to explore the robustness of the results.
Deviations of the images from the best fit models are also quantified. In
section \ref{sec:cap4} we study the distribution of derived parameters for
those galaxies which could be successfully fitted. We also identify and
correct for a serious (and previously known) systematic which results in the
assignment of significant disks to bright galaxies where disks are in most
cases actually absent. This section presents our principal results, estimates
of the fractions of the total light of all galaxies of given luminosity which
are in disks, in bulges, in galaxies with no detectable disk or in galaxies
with no detectable bulge. Combining these with previous measurements of the
galaxy luminosity function allows us to obtain corresponding fractions for the
galaxy population as a whole. A final section discusses and summarises these
results.

Throughout this paper, unless otherwise stated, we assume a Hubble constant of
$H_{0}=70$ $km$ $s^{-1}$ $Mpc^{-1}$ and an $\Omega_{M} = 0.3$,
$\Omega_{\Lambda} = 0.7$ cosmology in calculating distances and luminosities.

\section{Observational data}
\label{sec:cap1}

\begin{table*}
\begin{tabular}{|l|c|c|c|c|c|c|c|c|c|}
\hline
 & \multicolumn{8}{c|}{\bf Subsamples} &\\
\hline
 & $0 \le T < 1$ & $1 \le T < 2$ & $2 \le T < 3$ &
$3 \le T < 4$ & $4 \le T < 5$ & $5 \le T < 6$ & $T = 6$ & $T = -1$ & Total \\
\raisebox{1.5 ex}[0pt]{\bf Sample}  & E & S0 & Sa & Sb & Sc & Sd & Irr & unclassified &\\
\hline
Photometric      & 487 & 417 & 313 & 312 & 232 & 48 & 25 & 28 & 1862 \\
Spectroscopic    & 413 & 363 & 272 & 262 & 197 & 42 & 16 & 23 & 1588 \\
\hline                                                                     
\end{tabular}
\caption[Visual classification of our photometric and spectroscopic samples 
into morphological subsamples]
{Visual classification of our photometric and spectroscopic samples 
into morphological subsamples.}
\label{tab:sample}
\end{table*}
\begin{figure}
  \resizebox{1.15\hsize}{!}{\includegraphics{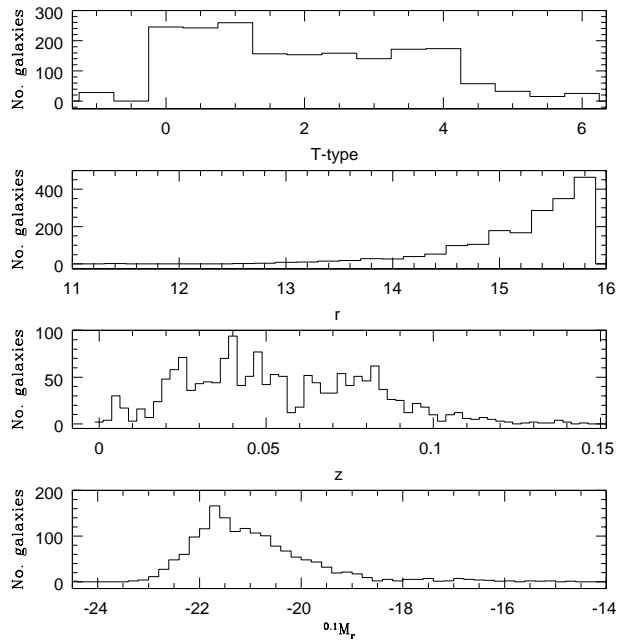}}
  \caption[The galaxies distribution in our sample versus morphological type,
  Petrosian magnitude in the $r$ band, redshift and $r$ band absolute
  magnitude] {The distributions of galaxies in morphological type (upper
  panel) and in extinction-corrected Petrosian apparent magnitude in the $r$
  band (second panel) are shown for the 1862 galaxies in our sample. The
  redshift (third panel) and corresponding absolute magnitude (lower panel)
  distributions are given for the 1550 galaxies in the sample for which we
  have spectroscopic information.}
\label{fig:thesample}
\end{figure}

\subsection{Galaxy sample}
In June 2001 the Sloan Digital Sky Survey (SDSS; \citet{York:00}) released its
Early Data Release (EDR;\citet{EDR:02}), roughly 462 square degrees of imaging
data collected in drift scan mode. 
The imaging is conducted in the $u-$, $g-$, $r-$, $i-$ and $z-$bands
(\citet{Fukugita:96}; \citet{Gunn:98}; \citet{Hogg:01}; \citet{Smith:02}; \citet{Pier:03}).
The reader is referred to \citet{Ivezic:04} for details on the phtometric quality assessment. 
The EDR contains around a million galaxies distributed
within a narrow strip of 2.5 degrees across the equator. As the strip crosses
the galactic plane, the data are divided into two separate sets in the North
and South Galactic caps.  The SDSS has the ambitious goal to image a quarter
of the Celestial Sphere at high Galactic latitude as well as to obtain spectra
uniformly for all the brighter galaxies. For the present project this has the
advantage, in comparison to previous work, of having uniform photometry and
spectroscopy over a much larger area, permitting a major improvement in sample
size and homogeneity.

In the following analysis we are using a sample of galaxies defined by the
Japanese Participation Group (JPG, \citet{Nakamura:03}). This is a homogeneous
sample obtained from the northern equatorial stripes of the SDSS EDR.  The
region of the sky covered is an area of 229.7 square degrees at
$145.15^{\circ} \le \alpha \le 235.97^{\circ}$ and $-1.27^{\circ} \le \delta
\le 1.27^{\circ}$.  The sample is limited to bright galaxies with $r \le 15.9$
after Galactic reddening correction. Eye classifications cannot be made
confidently beyond this magnitude, and for our current purpose this bright
limit has the advantage that the galaxies are all large compared to the SDSS
point-spread function. 

All the 1862 galaxies in the sample were classified by eye on the system of
the {\it{Hubble Atlas of Galaxies}} \citep{Sandage:61} by JPG scientists using
the $g$ band image of each galaxy. For each galaxy the final quoted type is
the mean of 4 independent classifications by different scientists. The {\it
rms} of these 4 classifications is also given; they typically agree within
$\Delta T \le 1.5$.  The corresponding numerical classification as defined in
the {\it{Third Reference Catalogue of Bright Galaxies}}
\citep{deVaucouleurs:91} is also reported.  The seven resulting subsamples
separate galaxies according to morphology going from ellipticals (E) through
lenticulars (S0) and early--type spirals (Sa, Sb), to late--type spirals (Sc,
Sd) and irregulars (Irr).  For 1588 galaxies out of our sample of 1862 we have
spectroscopic information.  A summary of the distribution of our galaxies
across the morphological classes is given in Table \ref{tab:sample}.  We refer
to \citep{Nakamura:03} for further details.  Further work exploring the
properties of this sample can be found in \citet{Nakamura:04}, 
\citet{Kelly:04} and \citet{Fukugita:04}.

\subsection{Photometric and spectroscopic data}

Two important quantities used in this paper are taken directly from the SDSS
database: the redshift and the Petrosian magnitude.  
The first is obtained by the spectroscopic pipelines {\bf idllspec2d} 
(written by D.Schlegel \& S.Burles) and {\bf spectro1d} (written by M. 
SubbaRao, M. Bernardi and J. Frieman).
A description of the tiling algorithm used to assign targets to each pointing
is given in \citet{Blanton:03c}. The SDSS spectroscopic galaxy samples consist
of all galaxies brighter than $r=17.77$ \citep{Strauss:02} and of a sample of 
luminous red galaxies \citep{Eisenstein:01} extending at $r<19.2$.
The distribution of galaxies with respect to $z$ for our sample is shown in
the third panel of Figure \ref{fig:thesample}.  
The second quantity is obtained by the
{\bf Photo} pipeline (see \citet{Lupton:01,Lupton:02}) and it is based on a
modified form of the Petrosian system for galaxy photometry which is designed
to measure a constant (and large) fraction of the total light of a galaxy
independent of its characteristic surface brightness.  Three related
quantities also used in this paper are $R_{50}$ and $R_{90}$, defined as the
radii which include respectively 50 and 90 percent of the Petrosian flux in
the r band (see \citet{EDR:02}), and the concentration index 
$c \equiv R_{90}/ R_{50}$. The second
panel of Figure \ref{fig:thesample} shows the distribution of galaxies with
respect to their Petrosian magnitude after correction for foreground Galactic
extinction using the reddening map of \citet{Schlegel:98}. Such
extinction--corrected Petrosian magnitudes are used throughout this
paper. 

\section{Image analysis}
\label{sec:cap2}

\begin{figure*}
 \resizebox{0.9\hsize}{!}{\includegraphics{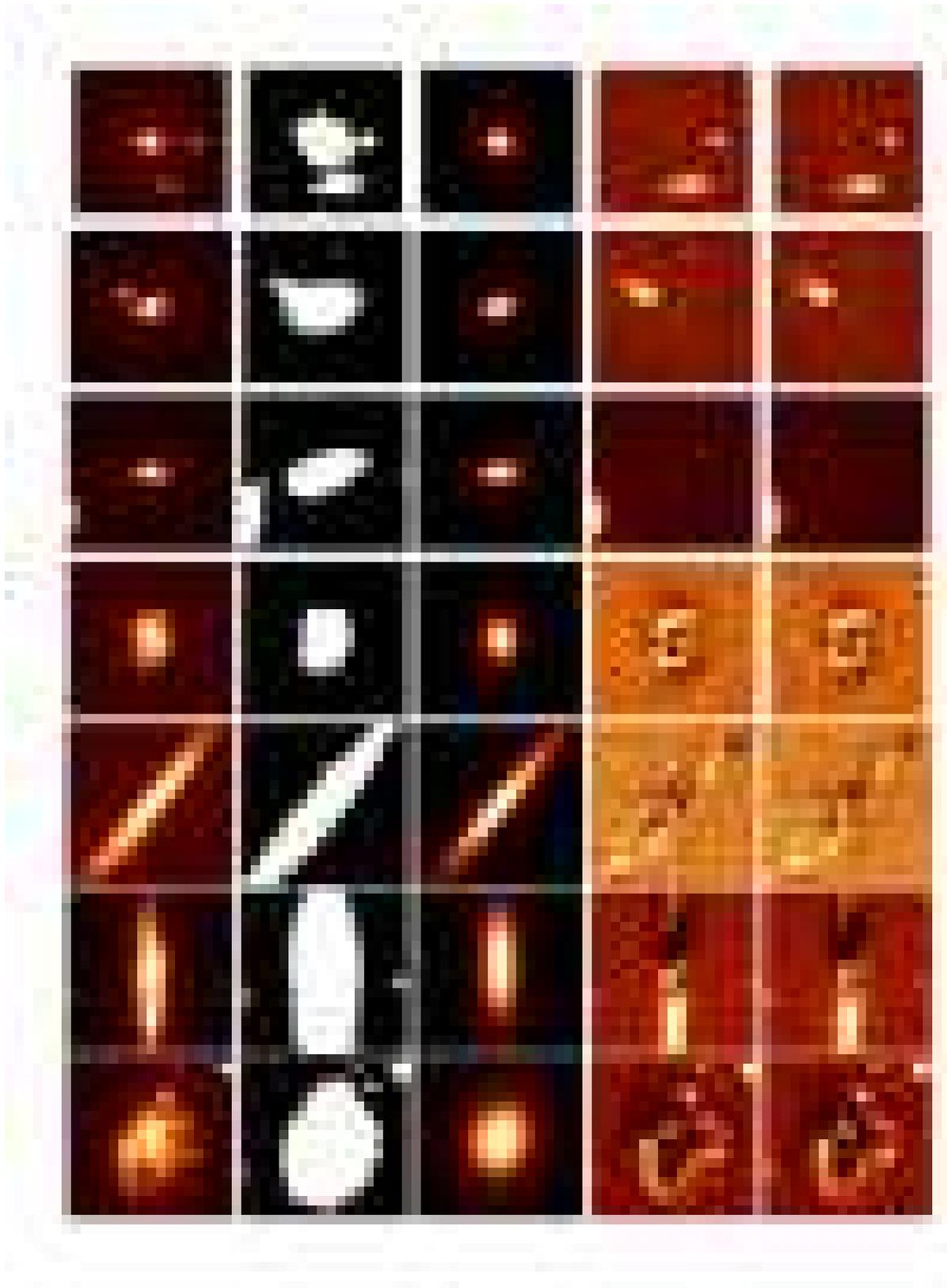}}
 \caption[Examples of science image, mask, model and residual images for 
  galaxies of different morphological types]
  {Examples of the science image, the mask, the model and the residual images 
  for our two fits (de Vaucouleurs plus exponential or S\'{e}rsic plus 
  exponential) in the $r$ band (from left to right)  
  for galaxies from our seven morphological classes: E, S0, Sa, Sb, Sc, Sd, 
  Irr (from top to bottom).}
\label{fig:image}  
\end{figure*}

\subsection{The fitting algorithm}
We examine the structural properties of our galaxies using {\bf Gim2D}
\citep{Simard:02}, a two--dimensional photometric decomposition algorithm
which fits each image to a superposition of an elliptical component with a
S\'{e}rsic profile, representing the bulge, and a concentric elliptical
component with an exponential profile, representing the disk.  It is important
to recognise that this separation into bulge and disk is based only on the
observed image, and may not correspond to the ``best'' decomposition if
additional information, for example from kinematics of the stars or the gas,
is taken into account.

It is well known that the bulge component of most galaxies can be well
represented by a surface brightness profile of S\'{e}rsic form:

\begin{eqnarray}
\Sigma(r)=\Sigma_{e} \cdot \exp{\{-b \, [(r/r_{e})^{1/n}-1]\}}
\label{eq:sersic}
\end{eqnarray}
where $\Sigma(r)$ is the surface brightness a distance $r$ from the centre
along the semi-major axis and $\Sigma_{e}$ is its characteristic value, the
effective surface brightness, defined as the value at the effective radius
$r_e$. The parameter $b$ is related to the S\'{e}rsic index $n$ and is set
equal to $1.9992 n - 0.3271$ so that $r_{e}$ is the projected radius enclosing
half of the total light \citep{Sersic:68, Ciotti:91}.  Many authors fit bulges
and ellipticals with a more specific function, the de Vaucouleurs $r^{1/4}$
law, which is obtained by setting $n=4$. When fitting S\'{e}rsic profiles in
the following we will assume $0.2<n<4$.  This choice is driven by the
knowledge that fits to low--luminosity ellipticals and bulges generally give
$n$ significantly smaller than 4, the standard value for bright ellipticals
(de Jong 1996; Caon et al.2005). Values of $n$ in excess of 4 are sometimes found for cD
galaxies but our sample does not contain such exceptionally luminous systems.

Disks are generally well described by an exponential profile (corresponding to
$n=1$), although non-axisymmetric features due to bars, spiral arms and dust
lanes can be large. We use the standard parametrisation:

\begin{eqnarray}
\Sigma(r)=\Sigma_{0} \cdot \exp(-r/h)\ ,
\label{eq:exp}
\end{eqnarray}
where $\Sigma_{0}$ is the central surface brightness and $h$ the
disk scalelength.

These laws are purely empirical fitting functions with no strong theoretical
justification. Other functions might provide equally good fits to a galaxy
profile leading to a different parametrisation for the bulge--to--disk
decomposition, and possibly different values for global parameters such
as the bulge-to-total luminosity ratio $B/T$.

With this model there is a maximum of twelve parameters which are adjusted in
fitting the galaxy image and that we retrieve as output from our
decomposition: the total flux of the object; the bulge--to--total light ratio
$B/T$, defined as the fraction of the total flux in the bulge component so
that $B/T=1$ corresponds to a pure bulge and $B/T=0$ to a pure disk; the bulge
effective radius $r_e$; the disk scalelength $h$; the disk inclination angle
$i$ defined so that $i=0$ for face--on disks and $i=90$ for edge--on ones --
the disk axial ratio is then $(b/a)_{disk}=cos(i)$; the bulge ellipticity $e$
given in terms of the bulge axial ratio by $e = 1 - (b/a)$; the bulge and disk
position angles (hereafter PA) measured clockwise from north and allowed to be
different; the S\'{e}rsic index $n$ which we sometimes fix at $n=4$; the x-y
pixel shifts $dx$ and $dy$ of the galaxy centre position
in the model and science thumbnail images; and the background intensity level.

Additional parameters could be introduced to model other features (e.g bars,
spiral arms, etc.) but such decompositions become somewhat arbitrary and may
not converge to unique solutions. Our current choice is standard and we found
it to be a good compromise between stability of results and flexibility of
representation. Notice that for an axisymmetric galaxy the position angles of
the bulge and the disk would be the same. Allowing them to differ makes it
possible for the code to detect triaxial bulges or bars in suitably oriented
galaxies.

\subsection{Image reduction}  
We perform our analysis starting from corrected frames of area $13^{\prime}.52
\times 8^{\prime}.98$ taken directly from the SDSS archive.  In these
large-scale images flat--field, bias, cosmic--ray, and pixel--defect
corrections have already been applied. The pixel-size in these images is 
0,396 arcsec. 
To proceed with our fitting, we begin with a list of source positions and
apply the {\bf SExtractor} galaxy photometry package version 2.2.2
\citep{Bertin:96} to each field to estimate the local sky background level at
each point and to define the isophotal area where each object is above the
detection threshold (we choose a threshold which is higher than background by
1.5 times the background noise).  When SExtractor performs galaxy photometry,
it constructs a segmentation (or mask) image in which pixels belonging to the
same object all have the same value and sky background pixels are flagged by
zeros. Our 2--D image fit is carried out on all pixels belonging to the same
SExtractor-defined object.  In practice {\bf Gim2D} uses thumbnail images, two
for each galaxy, extracted around the object of interest.  The area of these
is chosen to be 10 times larger than the mask area defined by SExtractor.  The
first thumbnail is cut from the corrected frame and is corrected for the local
background estimated by SExtractor, while the second contains the
corresponding pixels from the mask image. The fit then produces values and
uncertainty ranges for each model parameter. Subtraction of the best-fit
model from the input thumbnail produces a residual image that can be used to
characterise how well the model fits the galaxy. Examples of all these
images for representative galaxies from each of our morphological
classes are shown in Figure \ref{fig:image}.

When the fitting algorithm starts to sample the 12-dimensional parameter
space, it considers not only the pixels assigned to the main galaxy by the
mask but all pixels flagged as object or background in the SExtractor 
segmentation image. Important information about the galaxy could be contained 
in the pixels below the detection threshold.
In the residual images (see Figure \ref{fig:image}) one can see that 
all pixels belonging to other 
objects in the vicinity of the one under consideration are masked out.
The final flux is obtained by the integration of the best fit model over
all pixels, assuring that we do not lose the flux in the masked regions.

The model image of each galaxy is convolved with a point spread function (PSF) 
before comparison with the real data.  
The PSF can be highly variable across a corrected frame \citep{EDR:02}
and for this reason it is important to interpolate the PSF parameters measured
for individual stars to the position of each galaxy before convolving with the
model.  Our galaxy light model is thus the sum of an exponential disk and a
S\'{e}rsic, or de Vaucouleurs, bulge, convolved with this ``best'' PSF.

Finally a photometric calibration and the redshift are required to retrieve
physical quantities from the output of the B/D decomposition code. 

\section{Are we confident of our decompositions?}
\label{sec:cap3}

We have carried out the above fitting procedure entirely independently for
each of our galaxies in the $r$ and the $i$ bands. The two segmentation images
differ slightly and in addition there are colour variations across many of our
galaxies as a result of variations in the underlying stellar populations and
in the dust distribution. It is thus reassuring that the structural parameters
in the two bands are in good agreement in the great majority of cases (see
below). This shows that the fitting procedure produces stable results. In
addition {\bf Gim2D} produces acceptable converged parameter sets for almost
all the galaxies in both bands. In Table \ref{tab:modelled} we show the
number of galaxies from our spectroscopic sample that are successfully
modelled by the code in each band in the two cases when the S\'{e}rsic index
is set equal to 4 and when it is allowed to float. We consider a fit to
be successful when the code is converging.
As expected slightly more galaxies can be fit when $n$ is kept free, 
and in this case only $\sim 50$ of our 1834 galaxies
cannot be fit acceptably in either of the two bands. 
These galaxies are almost all later type spirals (Sb--Sc--Sd) which are not 
modelled simply because the centroid position provided by the SDSS database
does not match the one obtained with SExtractor within the defined 4 arcsec 
searching radius.
It is interesting to notice that the modelling also fails occasionally
for early--type objects when these are forced to follow
a de Vaucolueurs law for the central photometric component.
We now discuss aspects of these fits in more detail. 

\subsection{Comparison between fits}
\begin{figure}
 \resizebox{1.15\hsize}{!}{\includegraphics{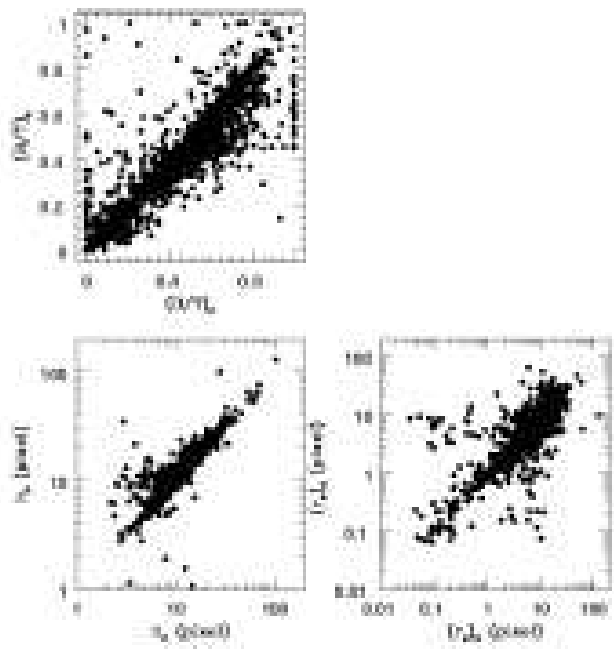}}
  \caption[Relation between bulge--to--disk ratio, disk 
  scalelength and effective radius of the bulge in the $i$ band
  retrieved using a de Vaucouleurs plus exponential and a S\'{e}rsic plus exponential fit.]
  {The bulge fraction (upper-left), the disk scalelength (lower-left) and the effective radius 
  of the bulge (lower-right) obtained using different parametric functions  
  to perform the decomposition (de Vaucoleurs plus exponential on the x-axis and S\'{e}rsic 
  plus expenential on the y-axis) are plotted against each other for 1702 galaxies in the
  $i$ band.}
\label{fig:diff_fits}  
\end{figure}

As already noted, we have fitted all our galaxies with a bulge model
in which the S\'{e}rsic index $n$ is free and also with a model in which
it is fixed to the de Vaucouleurs value $n=4$. In this subsection we
show that the parameters of the decomposition of most interest to
us are only weakly affected by this choice for most galaxies.
In Figure \ref{fig:diff_fits} we compare the values of the bulge fraction $B/T$, of
the disk scale length $h$ and of the bulge effective radius $r_e$
obtained for each galaxy in the two cases. We show results for
the $r$ band only (results for $i$ are similar).

\subsection{Comparison between different bands}
\begin{figure}
 \resizebox{\hsize}{!}{\includegraphics{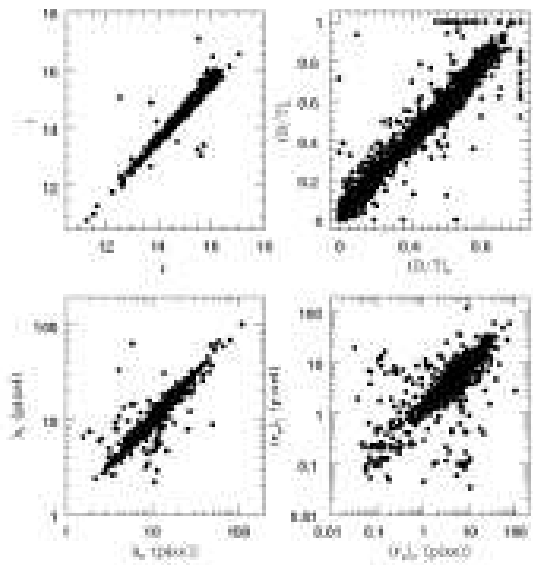}}
  \caption[Relation between total flux, bulge--to--disk ratio, disk 
  scalelength and effective radius of the bulge in the $i$ and $r$ bands
  retrieved using a de Vaucouleurs plus exponential fit]
  {The total flux (upper--left), the bulge fraction
  (upper--right), the disk scalelength (lower--left) and the effective
  radius of the bulge (lower--right) in the $i$ and $r$ bands are plotted 
  against each other for the 1638 galaxies modelled in both bands by fitting 
  a de Vaucouleurs profile to the bulge and an exponential to the disk.}
\label{fig:Gim2ddeVexp1}  
\end{figure}

\begin{figure}
 
\resizebox{\hsize}{!}{\includegraphics{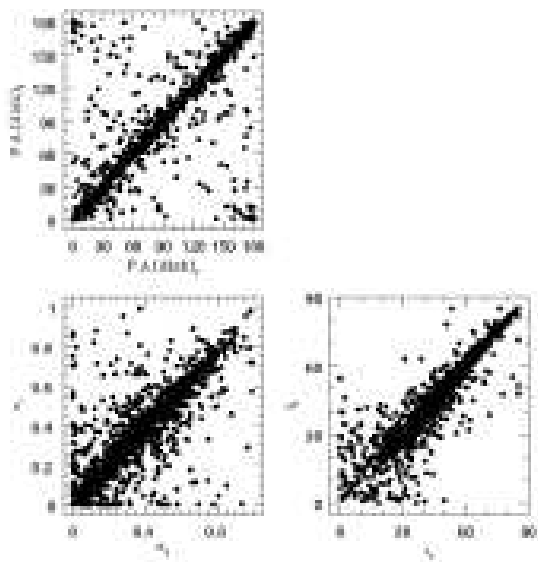}}
  \caption[Relation between position angle of the disk, 
  ellipticity of the bulge and inclination angle of the disk in the $i$ 
  and $r$ bands retrieved using a de Vaucouleurs plus exponential fit]
  {The position angle of the disk (upper--left), the ellipticity of the 
  bulge (lower--left) and the
  inclination angle of the disk (lower--right) in the $i$ and $r$ bands are 
  plotted against each other for the 1638 galaxies modelled in both 
  bands by fitting a de Vaucouleurs profile to the bulge and an exponential 
  to the disk.}
\label{fig:Gim2ddeVexp2}
\end{figure}

\begin{figure}
 \resizebox{\hsize}{!}{\includegraphics{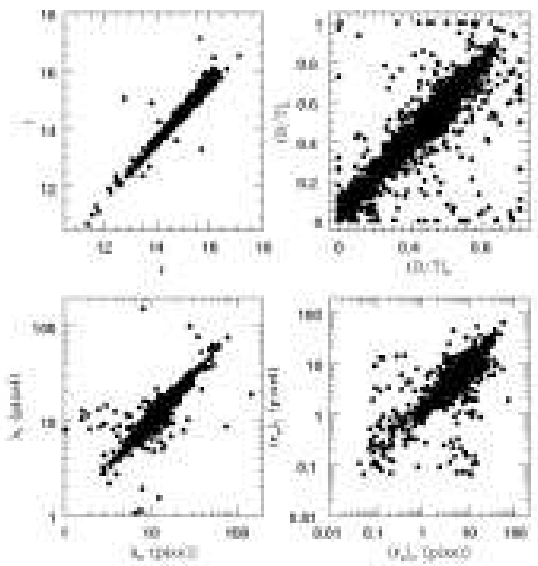}}
  \caption[The same as Figure \ref{fig:Gim2ddeVexp1} but for a S\'{e}rsic
  plus exponential fit]
  {The same as Figure \ref{fig:Gim2ddeVexp1} but for the 1766
  galaxies modelled in both bands by fitting a S\'{e}rsic profile to the
  bulge and an exponential to the disk.}
\label{fig:Gim2dsersicexp1}
\end{figure}

\begin{figure}

\resizebox{\hsize}{!}{\includegraphics{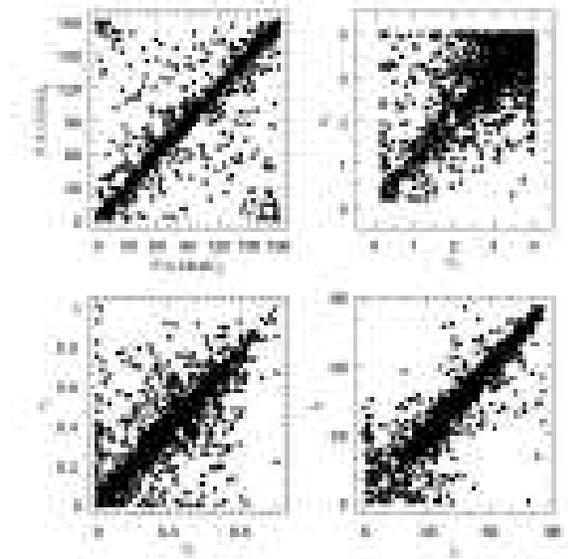}}
  \caption[Similar to Figure \ref{fig:Gim2ddeVexp2} but for a S\'{e}rsic
  plus exponential fit]
  {Similar to Figure \ref{fig:Gim2ddeVexp2} but for the 1766
  galaxies modelled in both bands by fitting a S\'{e}rsic profile to the
  bulge and an exponential to the disk. In this figure we also compare
  the values of the S\'{e}rsic index $n$ found in the two bands (upper right).}
\label{fig:Gim2dsersicexp2}
\end{figure}

In this section we compare the parameters estimated for each galaxy when the
same model is fit independently to images in each of the two SDSS bands
analysed here.  In Figures \ref{fig:Gim2ddeVexp1} and \ref{fig:Gim2ddeVexp2}
we plot the retrieved parameters in the Sloan $i$ and $r$ bands against each
other for the 1636 galaxies modelled successfully in both bands assuming a de
Vaucouleurs profile for the bulge and an exponential for the disk.  The
corresponding results are shown in Figure \ref{fig:Gim2dsersicexp1} and
\ref{fig:Gim2dsersicexp2} for the 1766 galaxies modelled successfully in both
bands assuming a S\'{e}rsic profile for the bulge and an exponential for the
disk; in this case, however, we also compare the values of the index $n$ found
in the two cases.

In the apparent magnitude plot there are only few isolated points
while the rest correlate very well, the scatter
being consistent with the expected variation in mean colour. The scatter in
bulge--to--total light ratio is also gratifyingly small for the vast majority
of the galaxies, particularly when $n$ is fixed to the de Vaucouleurs
value. The scatter is slightly larger when $n$ floats because different best
fit values of the index in the two bands (see Figure
\ref{fig:Gim2dsersicexp2}) lead to different splits of the luminosity between
bulge and disk. In most cases, however, similar $n$ values are found in the
two bands. The ``arrow'' shape of the $B/T$ plots reflects the fact that the
data in one band occasionally prefer a weak disk while no disk is present in
the best fit in the other band.

There is also quite good agreement between the values of the scalelengths for
the bulge and disk components measured in the two bands. The agreement is
worse for bulges than for disks and for components of small angular size
compared to larger ones. This is presumably a reflection of resolution
problems due to the finite pixel size and to difficulties with the PSF
deconvolution. Nevertheless the apparent axial ratios of both bulges and disks
agree well in the two bands with the scatter increasing for rounder
systems. 
The agreement between the position angle of the disk in the considered photometric 
bands is pretty good even if a substantial scatter is present. 
The points in the two extreme corners could be moved by a simple rotation 
of $180^{\circ}$. In addition we check that the distribution of the position angle 
is consistent with a random orientation on the sky.


\subsection{Error estimates and goodness of fit}
\label{sec:good_fit}
\begin{figure}
 \resizebox{1.0\hsize}{!}{\includegraphics{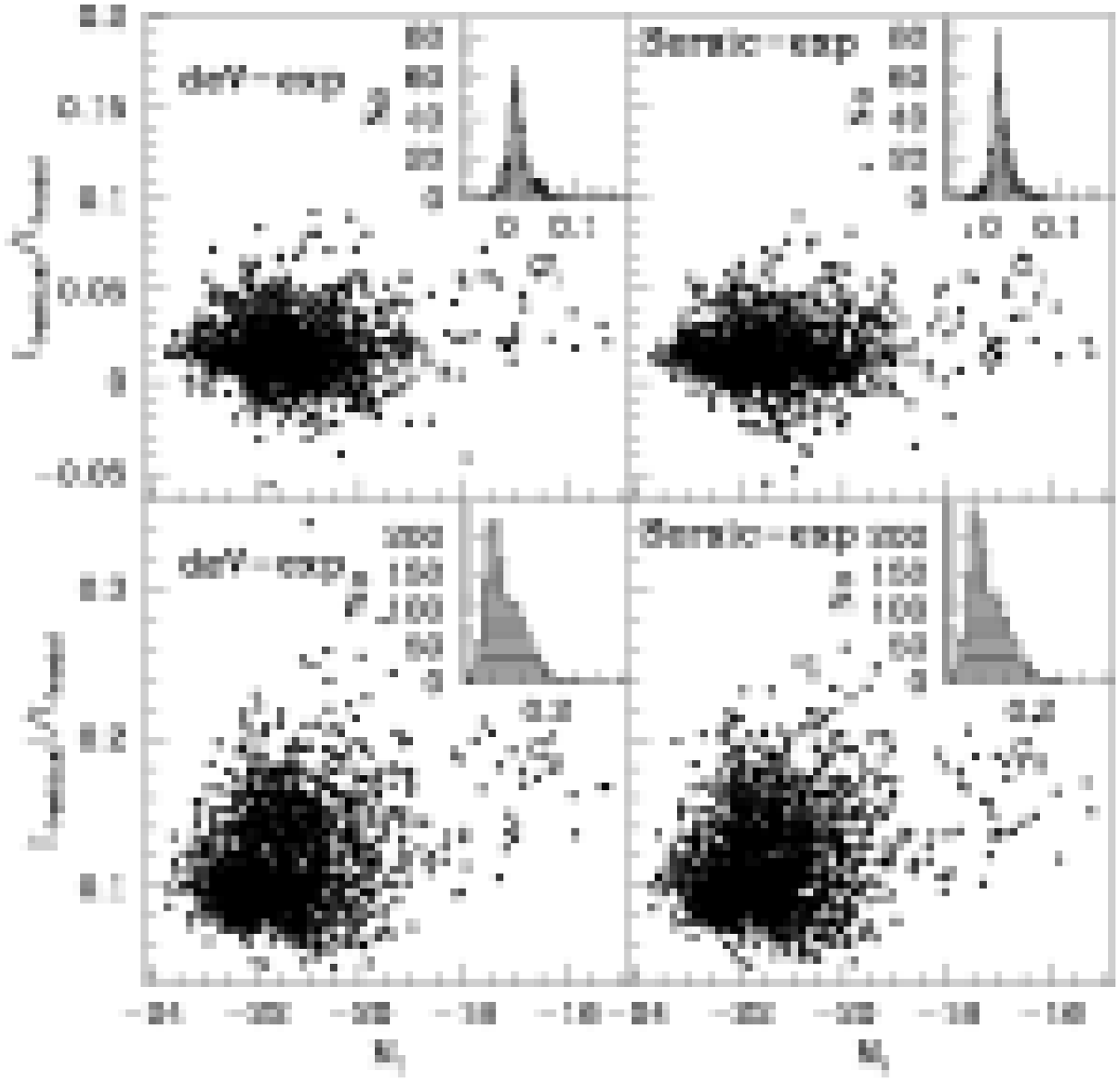}}
  \caption[Fraction of light in the residual image with respect to
  the total flux in the model image against the total
  absolute magnitude of the galaxy in the $i$ band]
  {Goodness of fit measures for our two-component modelling of the
  $i$-band images. The two left-hand panels give results for the 1469  
  galaxies modelled successfully using a de Vaucouleurs profile for
  the bulge and an exponential profile for the disk. The two right-hand
  panels give results for the 1532 galaxies modelled successfully when
  we use the more general S\'{e}rsic profile for the
  bulge. The two upper panels show results for $G_1$, the total light
  in the residual image in units of the total light of the model. The
  two lower panels give results for $G_2$, the sum of the absolute
  values of the individual pixel deviations from the model again in
  units of the total model luminosity. $G_1$ and $G_2$ are plotted
  against absolute magnitude in the main part of each panel while a
  histogram of their marginal distribution is given in the inset.} 
 \label{fig:goodnessfit_all}  
\end{figure}

When carrying out its fitting {\bf Gim2D} constructs a $\chi^2$ value for each
PSF-convolved model by summing over all pixels within the mask 
the square of the difference between model and data divided by
the variance of the pixel noise, assumed to be due entirely to photon
statistics. This measure of goodness of fit is then minimised over all
parameters (each required to lie within a prespecified ``allowed'' range) to
locate the maximum likelihood model.  Once the algorithm has converged, the
region of parameter space surrounding the likelihood maximum is sampled in
order to compute marginalised {\it a posteriori} one-dimensional probability
distributions for each model parameter. These are then used to define best
parameter estimates, taken to be the medians of these distributions, and
99\% confidence ranges defined by their upper and lower 0.5\% points.
Nevertheless he computed $\chi^2$ turns outs to be not too sensitive
to whether problems occur in the decomposition (i.e. a wrong point spread
function) or the decomposition reliably describes the light distribution in
the galaxy. 

With the aim of better understanding the goodness of the fit of our models we
introduce two additional measures, $G_1$ and $G_2$, which characterise the
size of the residuals without reference either to the overall luminosity and
size scales of a galaxy or to the expected counting noise.  They are defined
using the region flagged as belonging to the galaxy in the segmentation image
generated by SExtractor. $G_1$ is the difference between the model and
observed luminosities in this region as a fraction of the model luminosity,
while $G_2$ is the ratio of the sum of the absolute values of the residuals in
all the pixels to the model luminosity. Working from
the individual pixels ($ij$), by definition the total counts in the science
image ($\sum O_{ij}$) and in the model image ($\sum M_{ij}$) are due to the
light of the galaxy plus a uniform sky ($\sum S_{ij}$), which is the same in
the two images. The total counts in the residual image simply reflect the
difference between the luminosity of the observed and modelled galaxy ($\sum
D_{ij}=\sum O_{ij}-\sum M_{ij}$).  Our definitions can consequently be
formulated as:

\begin{eqnarray}
G_1=\frac{\sum D_{ij}}{L_{model}} \qquad\mbox{and}\qquad 
G_2=\frac{\sum|D_{ij}|}{L_{model}}
\label{eq:goodnessofthefit}
\end{eqnarray}
Figure \ref{fig:goodnessfit_all}
shows the distribution of galaxies with respect to these quantities as a
function of galaxy absolute magnitude. 
Results are presented for fits to the $i$ band data
both for floating $n$ and for $n$ fixed to the de Vaucouleurs value. Results
for the $r$ band are very similar.  The $G_1$ parameter is narrowly
distributed around zero, with a slight bias towards positive values. Thus {\bf
Gim2D} underestimates slightly the luminosities of these large
galaxies, but typically by only a couple of percent. Misestimates by more
than 5\% are very rare. The distribution of $G_2$ peaks at 0.1 and is skew
with a longer tail towards higher values. 
There is a tendency for deviations from the models to be larger for
intrinsically fainter galaxies, particularly below about $M_i =
-20$. Residuals are only slightly reduced by the extra freedom involved in
allowing $n$ to vary because, as can be seen from the examples in Figure
\ref{fig:image}, the dominant residuals are often due to non-symmetric
features such as spiral arms or dust lanes. We are encouraged that $G_2$ is
less than 15\% for about three quarters of our galaxies and almost never rises
as high as 25\%. We therefore believe that our model represents the
images of the majority of galaxies adequately for our purpose, and that
derived parameters can be used meaningfully to characterise physical
properties of the galaxies themselves.

\section{Results}
\label{sec:cap4}
\subsection{Comparison of S\'{e}rsic index distributions}
\begin{figure}
 \resizebox{1.15\hsize}{!}{\includegraphics{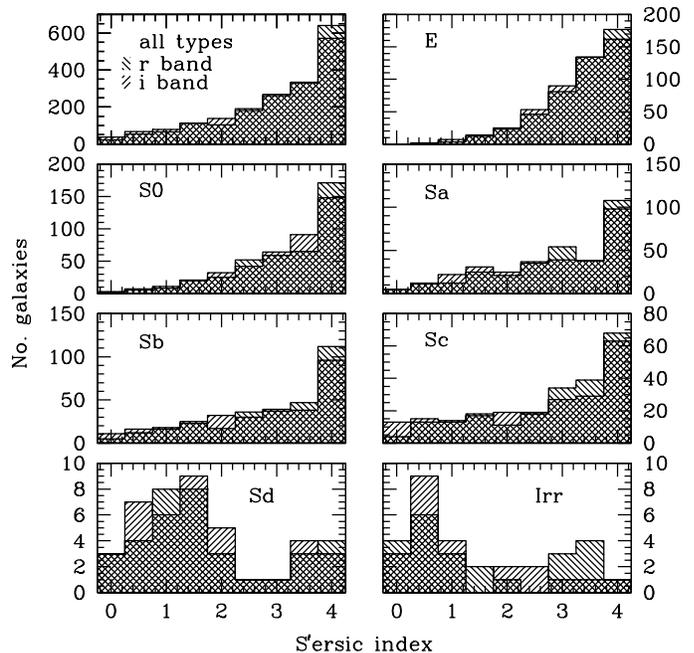}}
  \caption[Distribution of the S\'{e}rsic index $n$]
  {Distribution of the S\'{e}rsic index $n$. The upper--left
  panel shows the distribution for the total sample
  of 1790 and 1782 galaxies successfully modelled in the $i$ and $r$
  band respectively. In the other panels the S\'{e}rsic index
  distribution is separated according to morphological type.}
\label{fig:sersicindex}  
\end{figure}

Previous studies \citep{Andredakis:98, Courteau:96, deJong:96b}
have shown that the S\'{e}rsic formula can
fit the light profiles of many nearby ellipticals and bulges extremely well.
While massive systems usually require large values of $n$, similar to the de
Vaucouleurs value, less massive ellipticals and bulges, particularly the
bulges of late-type spirals usually demand smaller $n$ values and indeed can
often be well fit by an exponential law with $n=1$.  For our JPG sample when
we allow the S\'{e}rsic index to assume any value between 0.2 and 4 we obtain
the distributions shown in Figure \ref{fig:sersicindex}. The upper--left
histograms give results for the 1790 galaxies in the $i$ band and the 1782
galaxies in the $r$ band (out of the the total sample of 1862) which are
successfully fit by {\bf Gim2D}. These distributions peak at $n=4$, confirming
that the $r^{1/4}$ law provides an acceptable fit to the bulge component of a
large fraction of the galaxies in a magnitude-limited sample.  In the other
panels of Figure \ref{fig:sersicindex} we split the sample by morphological
type and it becomes evident that most E/S0 galaxies are well described by
values of $n$ close to than 4. The same is true for the bulges of most
early--type spirals.  For later--type spirals there is a clear shift to lower
values of $n$.  for many of these objects $n\sim 1$ is preferred, confirming
the earlier studies referred to above. Note that the distributions in 
 Figure \ref{fig:sersicindex} are almost independent of the band in which
the decomposition is carried out, confirming the robustness of the results.

\subsection{Disk and Bulge luminosity}
\label{sec:B/D_lum}
\begin{table*}
\begin{tabular}{|l|c|c|}
\hline
\multicolumn{3}{|c|}{\bf Galaxies modelled} \\
\hline
Type of fit & $r$ band & $i$ band \\
\hline
de Vaucouleurs + exponential  & 1450 & 1469  \\
S\'{e}rsic + exponential      & 1528 & 1532  \\
\hline                                                                     
\end{tabular}
\caption[Galaxies in the spectroscopic sample used to calculate
the fraction of light in the local universe in bulges and disks]
{Galaxies in the spectroscopic sample that we use to calculate
the fraction of light in the local universe in disks and bulges.
Of the 1588 galaxies in our sample for which the redshift is known,
1517 are successfully modelled by the code in both bands
when using the S\'{e}rsic profile and 1409 if we adopt the de 
Vaucouleurs law.
In the table we show
the number of galaxies in the $i$ and $r$ bands for which we have the
spectroscopy, the decomposition parameters and good photometry.}
\label{tab:modelled}
\end{table*}
\begin{figure}
\resizebox{1.15\hsize}{!}{\includegraphics{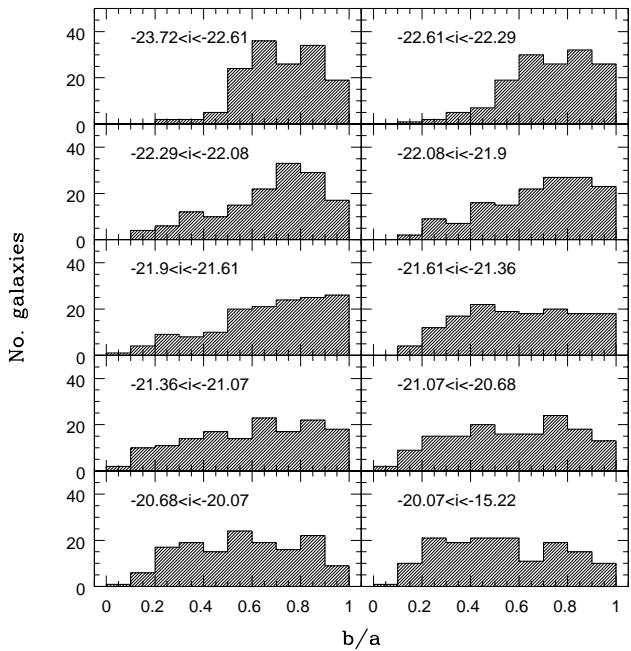}}
  \caption{Distribution of apparent axial ratio for the disk
  components of the 1469 galaxies with a
  redshift which were modelled successfully in the $i$ band by a de
  Vaucouleurs plus an exponential. 
  Galaxies are split by absolute magnitude into 10 bins containing
  approximately equal numbers of objects.}
\label{fig:b/a}  
\end{figure}
\begin{figure}
\resizebox{1.15\hsize}{!}{\includegraphics{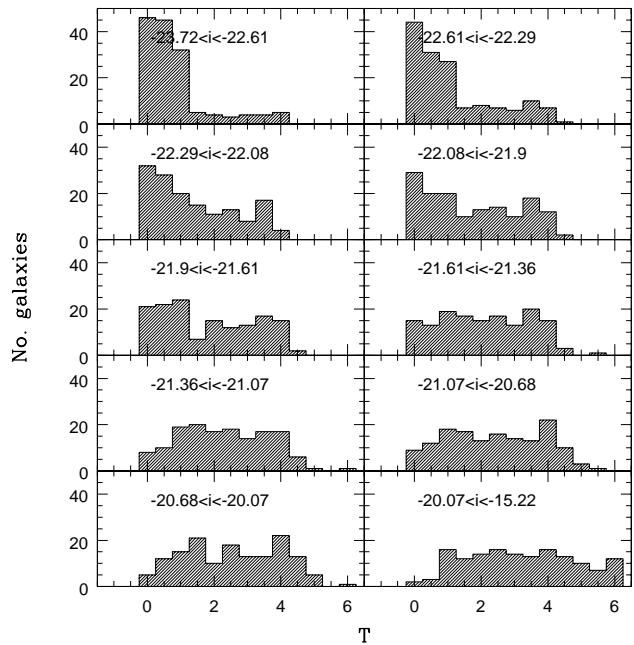}}
  \caption{As Figure \ref{fig:b/a} but showing the distribution over
 morphological type.}
\label{fig:T}  
\end{figure}
\begin{figure}
\resizebox{1.15\hsize}{!}{\includegraphics{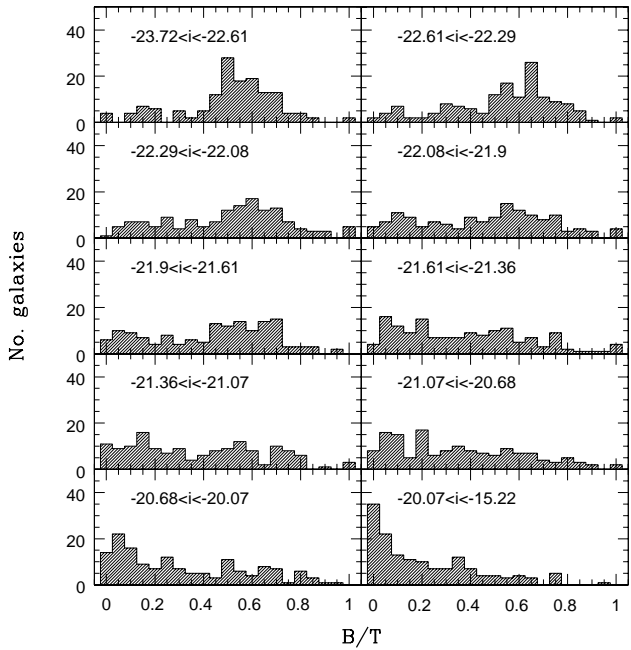}}
  \caption{As Figure \ref{fig:b/a} but showing the distribution over 
  bulge--to--disk ratio.}
\label{fig:BT}  
\end{figure}
\begin{table*}
\begin{tabular}{|l|c|c|}
\hline
Type of fit & $r$ band & $i$ band \\
\hline
de Vaucouleurs + exponential  & $(58.92 \pm 2.40) \%$ & $(54.92 \pm 2.02) \%$ \\
S\'{e}rsic + exponential      & $(54.82 \pm 1.95) \%$ & $(55.41 \pm 1.98) \%$ \\
\hline                                                                     
\end{tabular}
\caption[Total fraction of light in disks in the local
universe in the $r$ and $i$ bands]
{Total fraction of the light in disks in the local
universe in the $r$ and $i$ bands and for fits requiring $n=4$ and
allowing $n$ to vary over $0.2\leq n\leq 4$.}
\label{tab:frac_disk_light}
\end{table*}
\begin{table*}
\begin{tabular}{|l|c|c|}
\hline
Type of fit & $r$ band & $i$ band \\
\hline
de Vaucouleurs + exponential  & $(33.40 \pm 0.75) \%$ & $(31.15 \pm 0.73) \%$ \\
S\'{e}rsic + exponential      & $(34.39 \pm 0.72) \%$ & $(35.82 \pm 0.74) \%$ \\
\hline                                                                     
\end{tabular}
\caption[Total fraction of light in pure bulge systems in the local
universe in the $r$ and $i$ bands]
{Total fraction of the light in pure bulge systems in the local
universe in the $r$ and $i$ bands and for fits requiring $n=4$ and
allowing $n$ to vary over $0.2\leq n\leq 4$.}
\label{tab:frac_pure_bulge_light}
\end{table*}

In this section we derive the fraction of the luminosity density in the local
universe in bulges and in disks for the Sloan $i$ and $r$ bands, starting from
our complete sample of $r$--selected galaxies with $r<15.9$ (after correction
for Galactic extinction).  From our total sample of 1862 galaxies we here
consider only the 1588 objects for which spectroscopic data are available and
it is therefore possible to measure the absolute magnitudes needed by our
estimation procedure.  The absolute magnitudes used in this paper are
k--corrected using the code of \citet{Blanton:03}, $v2\_16$.  In the SDSS main
galaxy sample as a whole the median redshift is near $z=0.1$, so Blanton chose
to express results in the SDSS filter system shifted by 0.1. For our sample
the median redshift is about 0.05.  Nevertheless for consistency with other
SDSS work (in particular, with the luminosity functions we use below) we
follow Blanton's convention and k--correct to $z=0.1$. We denote absolute
magnitudes in this system as $^{0.1}M_r$ and $^{0.1}M_i$ to distinguish them
those in the unshifted system. Since our primary results concern the {\it
ratios} of luminosities in different components, this choice has no
effect on our analysis

There are 23 objects in this set for which no morphological type was assigned
by the JPG astronomers. We exclude these from further consideration here, 
leaving 1565 objects with redshift and a well defined ``by eye'' morphological
type.  From this sample we also excluded all galaxies for which
{\bf Gim2D} failed in the modelling, and in addition we removed
three objects for which the k--correction is not reliable due to bad
photometric data in the bluest and reddest bands.  This reduced the number of
objects used to estimate the luminosity densities in bulges and disks to the
numbers in Table \ref{tab:modelled}. 

Our strategy for computing the fraction of the local luminosity density which
is in bulges and disks is as follows. We separate the galaxies in our
sample into 11 bins according to their absolute luminosities in the $r$ or
$i$ bands. The nine brightest bins each contain about 10\% of the sample while
the two faintest bins contain about 5\% (we made this choice in order to get
better luminosity coverage for faint galaxies.)  For each bin we then use our
decompositions to estimate the fraction of the total light coming from disks
for galaxies at that absolute magnitude. Assuming this fraction to be
appropriate for all galaxies of similar intrinsic brightness, we combine it
with luminosity functions determined from much larger SDSS samples by
\citet{Blanton:03b} to obtain the fraction of the local luminosity density
which is in disks. The complementary fraction is then the amount in bulges.

While this appears straightforward, a serious complication arises from the
fact that 2-D fitting codes like {\bf Gim2D} tend to
fit radial variations in axial ratio or position
angle in ellipsoidal galaxies by assigning a fraction of their light to a
disk, when in fact none is present.  This systematic is well known and is
commented on in \citet{Simard:02}. We can demonstrate its presence in our
purely luminosity-selected data by examining the distribution of disk apparent
axial ratio returned by {\bf Gim2D}.  The distribution of $b/a$ is expected to
be uniform on $[0,1]$ for randomly oriented thin disks. Figure \ref{fig:b/a}
shows the distributions we actually obtain for the ``disks'' in our sample,
split into 10 equal bins by absolute total $i$-magnitude. While for faint
galaxies these distributions are indeed consistent with being flat, in the
brighter bins there is clearly a strong bias towards high b/a.  Among the
brightest galaxies almost no ``disks'' are found with $b/a<0.5$. Figure
\ref{fig:T} shows that these bright bins are dominated by early-type galaxies
according to the visual classifications of the JPG. The absence of small $b/a$
values demonstrates that few of these systems actually have significant thin
disks, despite the fact that {\bf Gim2D} assigns most of them $B/T$
ratios substantially smaller than unity (see Figure \ref{fig:BT}).

In order to circumvent this problem in the following analysis we use only the
galaxies in each absolute magnitude bin which have $b/a < 0.5$. We believe the
great majority of these must be true disks since ellipticals with apparent
axial ratios smaller than 0.5 are very rare. For random orientation the total
number of true disks expected in the bin is just twice the number with
$b/a<0.5$. The total light in disks in the bin is, however, more than twice
the light in the disks with $b/a<0.5$, since dust extinction is significantly
stronger in edge-on than in face-on disks. This must be corrected if we wish
to obtain an unbiased estimate of the luminosity density in disks.

In practice, our procedure works as follows. For each absolute magnitude bin
$k$ we estimate the fraction of the light in the disk component as
\begin{equation}
f_{disk,k}=\frac{L_{disk,k}}{L_{tot,k}}
\label{eq:fracbin}
\end{equation}
where $L_{disk,k}$ is the total luminosity of the disks
of the galaxies in the bin and $L_{tot,k}$ is the total luminosity 
from all components of these same galaxies. 
Since we assume that we can rely on our decomposition only for edge--on
systems, we split the numerator into two parts
\begin{equation}
f_{disk,k}=\frac{L_{b/a<0.5,disk,k}+L_{b/a>0.5,disk,k}}{L_{tot,k}}
\label{eq:fracbin2}
\end{equation}
where $L_{b/a<0.5,disk,k}$ is the luminosity due to ``edge-on''
disks with $b/a<0.5$ and is obtained directly from our decompositions. 
We estimate the luminosity $ L_{b/a>0.5,disk,k}$ in ``face-on'' disks
by assuming that true disks are randomly oriented and that their
internal extinction $A_{\lambda}$ depends on inclination according to the
standard prescription
\begin{equation}
A_{\lambda}=\gamma_{\lambda}\log(a/b).
\label{eq:extintion}
\end{equation}
Here $A_{\lambda}$ is the correction to exactly face--on orientation. 
Assuming this formula, it is a simple matter to relate the total
luminosity density in disks to that in disks with $b/a<0.5$. We
find
\begin{equation}
L_{disk,k} = 2^{1+0.4\gamma_{\lambda}}\cdot L_{b/a<0.5,disk,k},
\label{eq:diskcorr}
\end{equation}
hence
\begin{equation}
f_{disk,k}= 2^{1+0.4\gamma_{\lambda}}\cdot \frac{L_{b/a<0.5,disk,k}}{L_{tot,k}}.
\label{eq:fracbin3}
\end{equation}
The bulge to total ratio $B/T$ is one of the structural parameters returned by
{\bf Gim2D} for each galaxy. In combination with the total absolute Petrosian
magnitude (taken directly from the SDSS database) it allows us to estimate the
disk luminosity of each object contributing to $L_{b/a<0.5,disk,k}$. We
obtain $L_{tot,k}$ simply by summing the Petrosian luminosities of all
galaxies in the bin regardless of their $b/a$.  We take values for
$\gamma_{\lambda}$ from the work of Tully et al. (1998). 
Since the numerical coefficient on the right hand side of equation \ref{eq:fracbin3} 
is only slightly smaller at $i$ than at $r$ band, we assume the value 2.56 for 
both photometric bands.

\begin{figure}
 \resizebox{1.15\hsize}{!}{\includegraphics{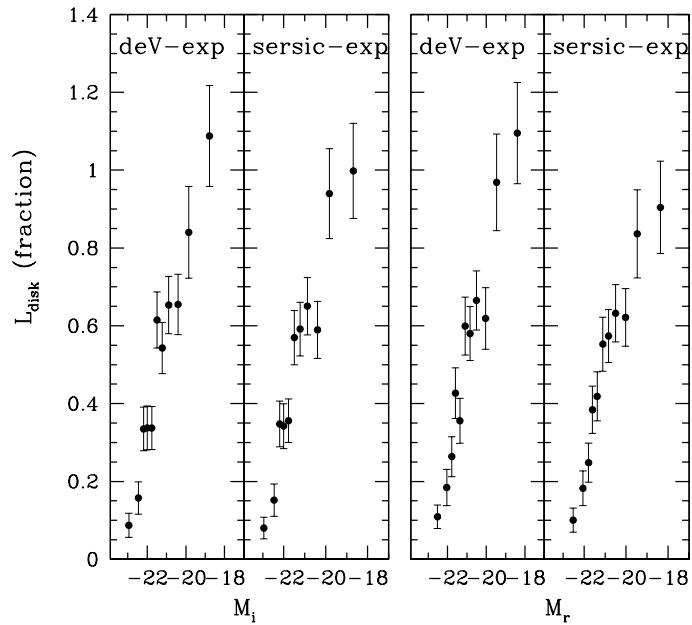}}
  \caption[Fraction of light in the galactic disk component as a function 
  of magnitude]
  {The two left--hand panels show the luminosity fraction in disks for galaxies in the Sloan $i$ band. 
  The galaxies are modelled with two different sets of parametric
  functions: a de Vaucoulours profile for the bulge and an exponential for the
  disk (left panel) or a S\'{e}rsic profile for the bulge plus an exponential 
  for the disk (right panel).
  The two right-hand panels show the same results but in the $r$ band.}
\label{fig:diskfrac}  
\end{figure}

The results of these calculations are shown in Figure \ref{fig:diskfrac}
for the galaxies modelled successfully in the $i$ and
$r$ bands. In each figure we show results separately for
decompositions which force a de Vaucouleurs bulge and for decompositions in
which we allow $0.2<n<4$.  The pattern is very similar in all cases. The
average light fraction in disks varies smoothly from about 10\% in the
brightest galaxies to almost 100\% in faint galaxies. For the brighter bins
these fractions are much smaller than the values obtained from a direct naive
average of the $B/T$ histograms of Figure \ref{fig:BT} because of the
systematic effect we have just been discussing.

The error bars on the points in Figure \ref{fig:diskfrac}
are important because they determine the precision of
our final results. We neglect the formal errors on $B/T$ returned by {\bf
  Gim2D} because these are quite small, typically $\pm 7 \%$, and are well
below the systematic error discussed above due to the assignment of isophote
twists or axial ratio changes to spurious thin disks. We assume,however, that
this systematic can be neglected for systems with $b/a < 0.5$. The uncertainty
in our estimate of the disk light fraction in each bin is then dominated by
sampling. As a statistical model for the population of a particular bin we
assume that a randomly chosen galaxy has a detectable edge-on disk ($b/a<0.5$)
with probability $p_{eo}$ where {\it a priori} we have $0<p_{eo}<0.5$. We also
assume that the $B/T$ values of these edge-on systems are drawn at random from
some unknown distribution with population mean and variance which we estimate
using the sample mean and variance, $\langle B/T\rangle$ and ${\rm Var}(B/T)$
respectively. If the bin contains $N_t$ galaxies of which $N_{eo}$ have disks
with $b/a<0.5$, then the maximum likelihood estimate of $p_{eo}$ is
$\tilde{p}_{eo} =N_{eo}/N_t$ provided $N_{eo}/N_t <0.5$ (it is equal to 0.5
otherwise). To approximate the variance of $\tilde{p}_{eo}$ we use the standard
binomial formula ${\rm Var}(\tilde{p}_{eo}) = \tilde{p}_{eo}(1 -\tilde{p}_{eo})/N_t$, even
though this is formally incorrect for $\tilde{p}_{eo} \sim 0.5$. Our estimate
of the mean light fraction in edge-on disks is then $(1-\langle
B/T\rangle)\tilde{p}_{eo}$ and we calculate the variance in this estimate as
$(\tilde{p}_{eo})^2{\rm Var}(B/T) + \langle B/T\rangle^2{\rm Var}(\tilde{p}_{eo})$.
These give the final results plotted when multiplied by the correction
factor of equation \ref{eq:diskcorr} which accounts for the light in face-on
disks. There is undoubtedly a systematic uncertainty associated with this
last step, but we ignore it here.

We can now average the disk light fractions of Figure  \ref{fig:diskfrac}
over the galaxy population as whole in order
to obtain the fractions of the total luminosity density in the
local universe coming from disks and from bulges.
The contribution of each of our absolute magnitude bins to the total
luminosity density $\Phi_{tot,k}$ can be obtained by integrating
the appropriate \citet{Blanton:03b} luminosity function across
the bin. The final result for the fraction of the local luminosity
density in disks is then, 
\begin{eqnarray}
f_{disk} =\frac{\sum_k \Phi_{tot,k} \cdot f_{disk,k}}{\sum_k \Phi_{tot,k}}.
\label{eq:dlumden}
\end{eqnarray}
Note that the luminosity functions for our sample are quite consistent with
those given by \citet{Blanton:03b}. We prefer to use the latter here
because of their much better statistical precision.  The final result we
obtain for the total fraction of the light coming from disks in the local
universe is $(54 \pm 2)\%$, with no detected dependence on observing band or
decomposition parametrisation.  The details are in Table
\ref{tab:frac_disk_light}. The error bars in this table are calculated
directly from those in  Figure \ref{fig:diskfrac}
assuming the uncertainties in the different
absolute magnitude bins to be independent. Uncertainties coming
from the luminosity function itself are negligible in comparison.

\begin{figure}
 \resizebox{1.15\hsize}{!}{\includegraphics{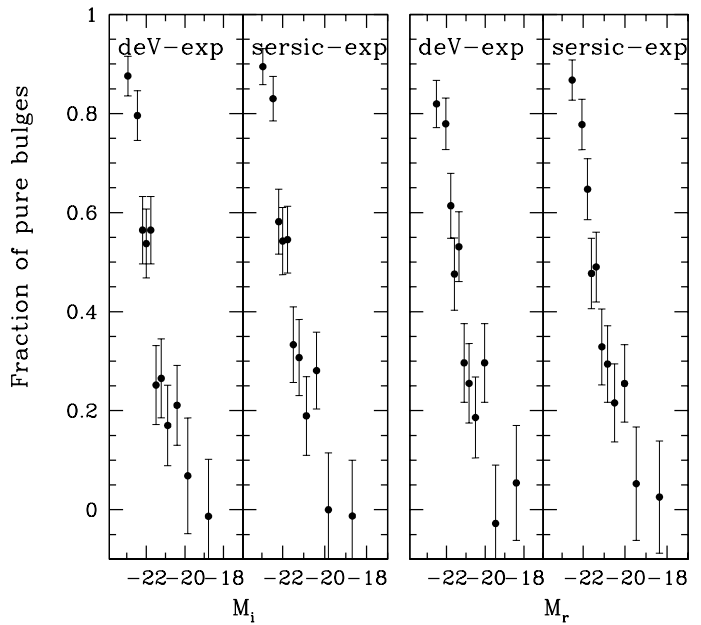}}
  \caption[Fraction of pure bulge galaxies as a function of the absolute
  magnitude]
  {The two left--hand panels show the fraction of pure bulge galaxies as a function of 
  the $i$ band absolute magnitude. 
  Different parametric functions are considered: de Vaucouleurs plus exponential (left--panel) 
  and S\'{e}rsic plus exponential (right--panel).
  The two right-hand panels show the same results but in the $r$ band.}
\label{fig:pure_bulges_frac}  
\end{figure}

A slight variation of this analysis allows us to calculate a second
interesting quantity: the fraction of galaxies in each of our absolute
magnitude bins which contain no detectable thin disk and so may be
considered ``pure'' bulge systems. Our hypothesis here is that {\bf Gim2D}
will detect any significant disk if it is sufficiently inclined to the 
line-of-sight that $b/a < 0.5$. Exactly one half of all disk galaxies
should be at least this inclined. We can thus estimate the number of
effectively diskless galaxies in each bin by subtracting twice the
number of objects with $b/a$ estimates below 0.5 from the total number
of galaxies in the bin. The result of this exercise is shown in Figure
\ref{fig:pure_bulges_frac} in 
similar format to Figure \ref{fig:diskfrac}.
Again the results are very similar in the two pass-bands and for our
two assumptions about bulge profiles. For the brightest bins we
find that the great majority of galaxies are effectively
diskless, while in the faintest bins our statistics are consistent
with at most a small fraction of ``pure bulge'' systems. The fraction
of ``pure bulges'' varies smoothly with absolute magnitude between
these two extremes. The error bars reflect the binomial uncertainty in our
estimate $\tilde{p}_{eo} =N_{eo}/N_t$ of the fraction of the bin population
with detectable edge-on disks. 

As before we can combine the results of Figure \ref{fig:pure_bulges_frac} 
with the luminosity functions of
\citet{Blanton:03b} to estimate the fraction of the luminosity density of the
local Universe which is contributed by effectively diskless systems.  The
result, given in detail in Table \ref{tab:frac_pure_bulge_light}, is about $(32 \pm
1)\%$ with no significant dependence on pass-band or bulge fitting
function. Thus the breakdown of stellar luminosity in the local Universe
is apparently 54\% in disks, 32\% in ``pure bulge'' systems with no
photometrically detectable (by {\bf Gim2d}) disk and only 14\% in the
bulges of galaxies with detectable disks. 

\section{Conclusions}
\label{sec:cap5}

We have used the two-dimensional photometric fitting programme {\bf Gim2D} of
\citet{Simard:02} on the images of an apparent magnitude limited sample of
1862 large galaxies selected from the SDSS and visually classified by
\citet{Nakamura:03}. In almost all cases the code returns a well-defined
decomposition of the galaxy into disk and bulge components with parameters
which have small formal error bars and vary little either between the $r$ and
$i$ band images analysed here or between decompositions in which the S\'{e}rsic
index of the bulge component varies or is fixed to the de Vaucouleurs value
$n=4$. The total amount of light in differences between the observed image and
the best fit model is typically only about 10\% of the galaxy luminosity.
Despite this apparent success, we show that for intrinsically bright galaxies
the ``disk'' component does not in most cases represent a true thin
axisymmetric disk, since the sample distribution of axial ratios $b/a$
deviates strongly from the uniform distribution expected for a population of
such disks -- near edge-on disks are grossly underrepresented. This problem
was already noted by \citet{Simard:02}. Apparently the code is using the
degrees of freedom provided by the assumed disk component to fit radial
changes in isophote shape or position angle within galaxies which have no
real thin disk.

We attempt to correct for this systematic problem by concentrating on galaxies
for which {\bf Gim2D} finds a disk axial ratio $b/a<0.5$.  We argue that
ellipsoidal stellar systems with such extreme apparent axial ratios are quite
rare so that the component isolated by the the decomposition programme is
likely to be a highly inclined disk. Since the selection criteria for our
sample depend at most weakly on inclination, we can use the assumption that
our galaxy sample is randomly oriented to correct from the sample with highly
inclined disks to the sample as a whole. Such corrections are at best
approximate, since there are undoubtedly cases where the decomposition mixes
light from the true bulge and disk components in such a way that
photometrically fit ``disk'' has $b/a>0.5$ even though the galaxy is
sufficiently inclined for the projected image of its true disk to have
$b/a<0.5$.  Without further information (for example, from kinematics) it is
very difficult to assess the size of the biases this may introduce.

As a first application we estimate average disk light fractions for galaxies
as a function of absolute magnitude. These range from about 10\% for the
brightest galaxies to almost 100\% for the faintest ones. At each absolute
magnitude we also estimate the fraction of ``pure bulge'' galaxies, defined as
galaxies for which {\bf Gim2D} would detect no disk with $b/a<0.5$ even if the
orientation were such that $b/a<0.5$ would be expected for any true
axisymmetric thin disk. We find that most of the galaxies in the brighter
absolute magnitude bins are ``pure bulge'' by this definition, but that this
fraction decreases steadily for fainter systems. We do not detect any
population of ``pure bulges'' in our faintest bin. These numbers differ
substantially from those inferred naively from the $B/T$ distributions
measured directly by {\bf Gim2D}: for example, in the brightest absolute
magnitude bin about 85\% of galaxies are ``pure bulge'' and disks only
contribute $\sim10\%$ of the total light, yet the median value of $B/T$
returned by the code is 0.55 and very few galaxies are fitted with $B/T>0.8$

By combining these results for the absolute magnitude dependence of the mean
light fractions in disks or in ``pure bulges'' with the luminosity functions
of \citet{Blanton:03b}, we have been able to estimate the fractions of all
galaxy light in the local Universe coming from disks and from ``pure bulges''.
The results depend little on the waveband used or on the bulge luminosity
profile we adopt. We find that $54\pm 2\%$ of the local luminosity density is
contributed by stars in disks and $32\pm 2\%$ by stars in ``pure bulges''. The
remaining $14\pm 3\%$ comes from bulges in systems with detectable disks. The
mean bulge-to-total ratio of the latter systems is thus $14/68\sim 0.2$,
substantially smaller than typical $B/T$ values in the histograms of
Figure \ref{fig:BT}. Note that only sampling uncertainties are taken into
account in the errors quoted here. Residual systematics may remain, reflecting
the fact that real galaxies have a more complex structure than the models we
use to describe them here. A fully convincing decomposition of galaxies into
bulge and disk components clearly cannot be performed using imaging data
alone. We believe, however, that the analysis of this paper provides the best
quantitative indication so far of the overall distribution of stars between
the two basic structural forms which make up galaxies.

\section*{acknowledgments}
We thank Luc Simard for help in using his excellent two-dimensional
fitting package {\bf Gim2D}.
Funding for the creation and distribution of the SDSS Archive has 
been provided by the Alfred P. Sloan Foundation, the Participating Institutions, 
the National Aeronautics and Space Administration, the National Science Foundation, 
the U.S. Department of Energy, the Japanese Monbukagakusho, and the Max Planck 
Society. The SDSS Web site is http://www.sdss.org/.

The SDSS is managed by the Astrophysical Research Consortium (ARC) for the 
Participating Institutions. The Participating Institutions are The University 
of Chicago, Fermilab, the Institute for Advanced Study, the Japan Participation 
Group, The Johns Hopkins University, the Korean Scientist Group, 
Los Alamos National Laboratory, the Max-Planck-Institute for Astronomy (MPIA), 
the Max-Planck-Institute for Astrophysics (MPA), New Mexico State University, 
University of Pittsburgh, University of Portsmouth, Princeton University, 
the United States Naval Observatory, and the University of Washington.

\addcontentsline{toc}{chapter}{Literaturverzeichnis}
   \bibliographystyle{mn2e}
   \bibliography{paperbib}
\clearpage

\label{lastpage}
\end{document}